\documentclass[twocolumn,reprint,prb,amsmath,amssymb]{revtex4-2}
\usepackage[utf8]{inputenc}
\usepackage{graphicx}
\usepackage{hyperref}

\begin{document}
%\MSMSE

\title[Structural phase-field crystal model for Lennard-Jones pair interaction potential]{Structural phase-field crystal model for Lennard-Jones pair interaction potential}

\author{V Ankudinov
%$^1$
%, N M Chtchelkatchev
%$^1$
%\footnote{Present address: Department of Physics, University of Bristol, Tyndalls Park Road, Bristol BS8 1TS, UK.} 
}

\address{Vereshchagin Institute of High Pressure Physics, Russian Academy of Sciences, 108840~Moscow (Troitsk), Russia}
%\address{$^2$ Department of Mathematics,  Imperial College, Prince Consort Road, London SW7~2BZ, UK}
%\address{$^3$ Department of Computer Science,  University College London, Gower Street, London WC1E~6BT, UK}
%\ead{vladimir@ankudinov.org}

\begin{abstract}
A modification of structural phase-field crystal (XPFC) model for an arbitrary pair interaction potential is presented. Formation of 1D and 2D structures  for the Lennard-Jones (LJ) potential  was studied numerically. The equilibrium lattice parameters for the presented structures were found consistent to the correspondent LJ-distance parameters. The lattice parameter of 2D triangle's structure matches the periodical in 1D, which shown to be consistent with the theory of freezing from the isotropic liquids. Numerically obtained XPFC phase diagram of two-dimensional structures qualitatively reproduces classical PFC diagram and coincides with the melting region of high-temperature part of LJ diagram.
\end{abstract}

\maketitle
\noindent{\it Keywords}: Phase-field crystal, structural phase-field crystal, pair interaction potentials, pair correlations, Lennard-Jones potential, Lennard-Jones potential
%\maketitle

\emph{ This is the version of the article before peer review or editing, as submitted by an author to 
``Modelling and Simulation in Materials Science and Engineering''. IOP Publishing Ltd is not responsible for any errors or omissions in this version of the manuscript or any version derived from it. The Version of Record is available online at \href{https://doi.org/10.1088/1361-651X/ac7e63}{doi:10.1088\/1361-651X\/ac7e63} }

\section{Introduction}

The phase-field crystal model (PFC) was formulated \cite{bib:pe,elder12,elder21,Greenwood2011,Toth2011} to describe continuous transitions from the homogeneous to the various periodic states (similarly to the Landau–Brazovskii transitions \cite{Landau1996,brazovskii75,bib:L}) and between them over diffusion times. The model is based on the description of a Helmholtz free energy, which is a functional of the atomic density field  which is periodic in the solid phase and homogeneous in the liquid (disordered) state. 
Recent advances in PFC-modeling of the different aspects of crystallization allow one to model many scenarios such as dynamics of freezing of colloids and polymers, epitaxial growth, ordering  on nano-scales~\cite{bib:adv2012,Granasy2019,VanTeeffelen2009} and  rapid crystallization~\cite{GDL}. Results of PFC simulations provide interface energies, pattern selection, multiphase solidification under non-equilibrium conditions, heteroepitaxy and multi-grain growth in presence of hydrodynamical flows~\cite{granasy,Elder2007,Guerdane2018,Fallah2013,Toth2010, Tang2011,Podmaniczky2017}. 
As a simplification of classical density functional theory (cDFT) of freezing~\cite{rama79,Ryzhov1979} the PFC model  utilizes several approximations~\cite{elder12,bib:berry2008} leading to the close to the hard-spheres-potential behavior of density peaks for a classical one-mode PFC. The validity of PFC was proofed by the molecular dynamics \cite{Guerdane2018,asadi14,Asadi2015a}. But at the same time, rough PFC approximations lead to a bit of criticism concerning about PFC unable to carefully predict surface energies on long times and inaccuracy in quantitative description of elastic properties and emerging structures~\cite{Baker2015,Archer2019}.

The form of the classical PFC free energy is close to the Swift-Hohenberg equation~\cite{ge-swift,bib:pe} and in general allows one to describe phase transitions of the first and second order \cite{Ankudinov20202}. Later to robust the structural transitions  the artificial multipeaked correlation kernels \cite{Greenwood2010,Greenwood2011} were introduced to the PFC models. Such kernels based on so-called structural PFC (XPFC) were generalized for a multi-component metallic systems \cite{Greenwood2011a,Ofori-Opoku2013,Fallah2013,Smith2017} and stable 3D structures such as diamond \cite{Chan2015,Baker2015}. Several cases of two-dimensional structures formation were also examined for three-point correlation functions~\cite{Seymour2016,Elder2018}.  Those improvements of XPFC model mostly regard to the modification of the position and size of first and second peak of correlation function in reciprocal space ($k$-space).     
  It was shown that stabilization of graphene-like structure requires not only a repulsive term but a rotationally invariant correlation function of higher approximation, which also capable to stabilize kagome lattice~\cite{mkhonta} (for the case of three-mode PFC).
 The multiscale PFC-like model were proposed by \cite{Zapolsky2017} where the excessive energy term was expanded in reciprocal space in two components: short ``condensation'' term and spherically symmetrical long-range interaction term. Obtained results clearly states that the repulsive term is sufficient for the ``structural'' crystallization in presence of middle-range attraction written in the form of derivative of $n^{th}$ mode. In \cite{Zapolsky2017} the quasi-continuous approximation of the phase field proposed for the case of discrete values (1 or 0) of the atomic density in each lattice node (about 8 nodes per lattice period).  
In cDFT the  expansion  of direct correlation functions  with quadratic terms  in reciprocal space  was carried out by \cite{Ghosh2017}. Such approach in fact is very close to the idea of eight-order \cite{jaatinen09} and twelve-order  \cite{Ankudinov2020a,Ankudinov2020b} fitting in PFC. Although the derivation of PFC as a consequent approximation of dynamical density functional theory with a gradient expansion involving derivatives leads to the possible instability above a certain value of the average density \cite{Archer2019}. The control of emerging in PFC structures is possible with the pair correlation function analysis proposed in \cite{Kondo2021}.
The promisable approach consisted in introduction a pair-correlation interactions  approximated by the rational function to XPFC is discussed in \cite{Pisutha-Arnond2013}.

In present work we propose a modification of the structural phase-field crystal (XPFC) model with the Lennard-Jones pair interaction potential. A study of the structure formation and crystallization from the undercooled liquid (homogeneous) phase was carried out.  In particular, the equilibrium lattice parameters, sequence of the emerged structures and their  phase diagram were studied.

\section{Classical density functional theory  of freezing and phase-field crystal model}
The classical PFC equation obtained from approximation of cDFT  is suitable for the description of the crystallization of liquids to the simple crystalline structures \cite{bib:adv2012,bib:pe, jaatinen09, wu07,Ryzhov1979}. 
The derivation of this model is based on the approximation of cDFT free energy of single-particle's probability density field $\rho({\bf {r}},t)$. 

Let one consider the static isothermal approximation of cDFT free energy containing ideal and excess free energies neglecting any external potential \cite{bib:pe,rama79,evans79,singh91}:
\begin{eqnarray}
\mathcal{F}[\rho( {\bf r} )] = \mathcal{F}_{id}[\rho( {\bf r} )]+\mathcal{F}_{ex}[\rho( {\bf r} )].
\label{FreeDFT-terms}
\end{eqnarray}
The  $\mathcal{F}_{id}$ contribution corresponds to the free energy functional of Boltzmann gas~\cite{evans79,bib:adv2012,bib:pe} and any particle's interactions were neglected:
\begin{equation}
\mathcal{F}_{id}([\rho({\bf {r}})]) = k_BT \int d{\bf {r}} \, \rho({\bf {r}})(\ln(\lambda^3 \rho({{\bf {r}}}) ) - 1),
\label{eq:free-DFT-id}
\end{equation}
where $k_B$ is the Boltzman constant,  $\lambda$ is the thermal de Broglie wavelength. Let one substitute normalized (by the reference density $\rho_0$) averaged atomic density field $n({\bf r})={\rho({\bf r})}/{\rho_0}-1$    to Eq.~(\ref{eq:free-DFT-id}):
\begin{equation}
\mathcal{F}_{id}([\rho({\bf {r}})])=
k_BT \rho_0\int\!\!d{\bf {r}} \, [(1+n({\bf {r}}))\,\,\mathrm{ln}\left( 1+n({\bf {r}})\right) -n({\bf {r}})].
\end{equation}

The second term $F_{ex}[\rho(\bf{r})]$ of the Eq.~(\ref{FreeDFT-terms}) corresponds to the excess energy of the particle's exchange interactions. To determine it explicitly, it is necessary to make a number of approximations \cite{evans79,singh91,Ryzhov1979}. 
The  exact expression of $F_{ex}[\rho({\bf r})]$ as a generating functional can be formally obtained with the expansion   around the small density change $\tilde{\rho}({\bf r})={\rho}({\bf r})-\rho_0$. Its expansion over time-independent (slowly varying) density field $\tilde{\rho}$ would be \cite{Evans1992,Hansen2013,rama79,Ryzhov1979}:
\begin{align}
\label{eq:spatial_expansion}
\mathcal{F}_{ex}&[\rho({\bf r})]=\mathcal{F}_{ex}[\rho_0]+\int d{\bf r} \tilde{\rho}({\bf r})
\frac{\delta \mathcal{F}_{ex}[\rho({\bf r})]  }{\delta \rho({\bf r})} + \nonumber \\
&\frac{1}{2} \int \int d{\bf r}  d{\bf r'} \tilde{\rho}({\bf r}) \tilde{\rho}({\bf r'})
\frac{\delta^2 \mathcal{F}_{ex}[\rho({\bf r})]  }{\delta \rho({\bf r}) \delta \rho({\bf r'})} + \\
&\frac{1}{6} \int \int d{\bf r}  d{\bf r'} d{\bf r''} \tilde{\rho}({\bf r}) \tilde{\rho}({\bf r'})\tilde{\rho}({\bf r''})
\frac{\delta^3 \mathcal{F}_{ex}[\rho({\bf r})]  }{\delta \rho({\bf r}) \delta \rho({\bf r'})\delta \rho({\bf r''})} + ... \nonumber
\end{align}
These functional derivatives of the excess energy are related to the $n$-body direct correlation functions $C^{(n)}$ \cite{Evans1992,Hansen2013}:
\begin{eqnarray}
\label{eq:pcf}
\frac{\delta^n \mathcal{F}_{ex}[\rho({\bf r})]  }{\delta \rho({\bf r}_1) \delta \rho({\bf r}_2) \cdot \cdot \cdot \delta \rho({\bf r}_n)} = -k_B T \, C^{(n)}({\bf r}_1,{\bf r}_2...{\bf r}_n).
\end{eqnarray}
Substituting the excess energy with direct correlation functions Eq.~(\ref{eq:spatial_expansion}) to Eq.~(\ref{FreeDFT-terms}), truncating it up to 2$^\mathrm{nd}$ term of Eq.~(\ref{eq:pcf}) and considering the normalized atomic density $n({\bf r})$ one can get a dimensionless free energy:
\begin{eqnarray}
F(n)=\frac{\mathcal{F}}{k_B T V {\rho}_0} = &
\int\!\!d{\bf {r}} \, [(1+n({\bf {r}}))\mathrm{ln}\left( 1+n({\bf {r}})\right) -n({\bf {r}})]- \nonumber \\
&\frac{1}{2}\int \!\!d{\bf {r}}\! \int \!\! d{\bf {r}}'\,\,n({\bf {r}}){C}\| {\bf {r}} -{\bf {r}}'\| n({\bf {r}}'),
\label{eq:pfc_der_0}
\end{eqnarray}
where  ${C}\| {\bf {r}} -{\bf {r}}'\|$ is the pair correlation function. 
Here the free energy is scaled by the energy of a reference state: $\mathcal{F}=\mathcal{F}[\rho( {\bf r} )] -\mathcal{F}[\rho_0] $. The pair correlation function could be approximated in  a reciprocal space (space of $k$-vectors) in a manner proposed by Ramakrishnan and Yusoff (RY)~\cite{rama79,Ryzhov1979} leading to the ``standard'' PFC-model~\cite{elder12,elder21}.

\subsection{Phase-field crystal}
 The ideal part $\mathcal{F}_{id}$ of the scaled free energy Eq.~(\ref{eq:pfc_der_0}) could be approximated in the form of Taylor expansion around the reference density $n({\bf r})=0$ \cite{elder12,elder21}. 
\begin{equation}
\mathcal{F}_{id}(n)=(1+n)\mathrm{ln}(1+n)-n\simeq \frac{a}{2}n^2
-\frac{b}{3}n^3
+\frac{v}{4}n^4.
\label{ldeg}
\end{equation}
Moreover such free energy introduces the first- and second-order phase transitions \cite{Ankudinov20202,jaatinen09}. The pair correlation function could be approximated in  a reciprocal space (space of $k$-vectors) in a manner proposed by Ramakrishnan and Yusoff (RY)~\cite{rama79,Ryzhov1979} as:
\begin{equation}
    \label{eq:gl_correlation_funcion_procedure}
{C^{(2)}(k)} \approx - C_0+C_2k^2-C_4k^4+...
\end{equation}
Thus, a truncated series of spatial derivatives in a direct space allow one to model the formation of periodic crystal:
\begin{equation}
\label{ckr}
\mathcal{F}_{ex}(n)= \frac{n}{2}(C_0 - C_2 \nabla^2 + C_4 \nabla^4 - ...)n
\end{equation}
The resulted classical PFC free energy read as
\begin{align}
{F}[n] =\!
\displaystyle \int\! \left[\frac{n}{2}{\cal L}n -
\frac{a}{3}n^3+\frac{v}{4}n^4\right]\! d{\bf {r}}, \\ {\cal L}_1 \equiv \Delta B_0 +B_0^x(q_{S0}^2+\nabla^2)^2.
\label{freen}
\end{align}
where ${\cal L}_1$ is the one-mode differential operator. 
 Proposed expansion could be truncated on higher modes introducing the two- or three-mode PFC models \cite{mkhonta,asadi15,zaeem16,Ankudinov2020a}.

\section{Structural PFC model for arbitrary interaction potentials}

\subsection{Structural PFC model}
This structural PFC (XPFC) utilizes a reduced expansion of pair correlation function in a row of exponents. The later development of XPFC  consists in the approximation of three-point correlation function  as a sum of pair-correlation functions which includes repulsive and anisotropic terms \cite{Seymour2016,Elder2018}. This artificial correlation functions allow one to reproduce stable hexagonal lattices, which also can be modeled with three-mode isotropic PFC model~\cite{mkhonta}.
%The excessive energy  $F_{ex}$ includes  the pair correlation functions in real space \cite{rama79,bib:adv2012}. In PFC model it is truncated on pair-correlation function:
%\begin{equation}
%\mathcal{F}_{ex}(n)=-\frac{1}{2}\int \!\!d{\bf {r}}\! \int \!\! d{\bf r}'\,\,n({\bf {r}}){C}_{2} \| {\bf {r}} -{\bf r}'\| n({\bf r}').
%\end{equation}
The XPFC free energy in dimensionless case could  be derived from Eq.~(\ref{eq:pfc_der_0}) with the Landau expansion Eq.~(\ref{ldeg}) as
\begin{eqnarray}
{{F(n)}} = 
\int&\!\!d{\bf {r}} \, \left[\frac{1-\varepsilon}{2}n^2
-\frac{a}{3}n^3
+\frac{v}{4}n^4\right]-  \nonumber \\
&\frac{1}{2}\int \!\!d{\bf {r}}\! \int \!\! d\mathbf{r'}\,\,n({\bf {r}}){C}_{2} \| {\bf {r}} -\mathbf{r'}\| n(\mathbf{r'}).
\label{eq:pfc_der_1}
\end{eqnarray}

\subsection{Correlation function for arbitrary interaction potentials}
\begin{figure}
(a)  \includegraphics[width=0.80\columnwidth]{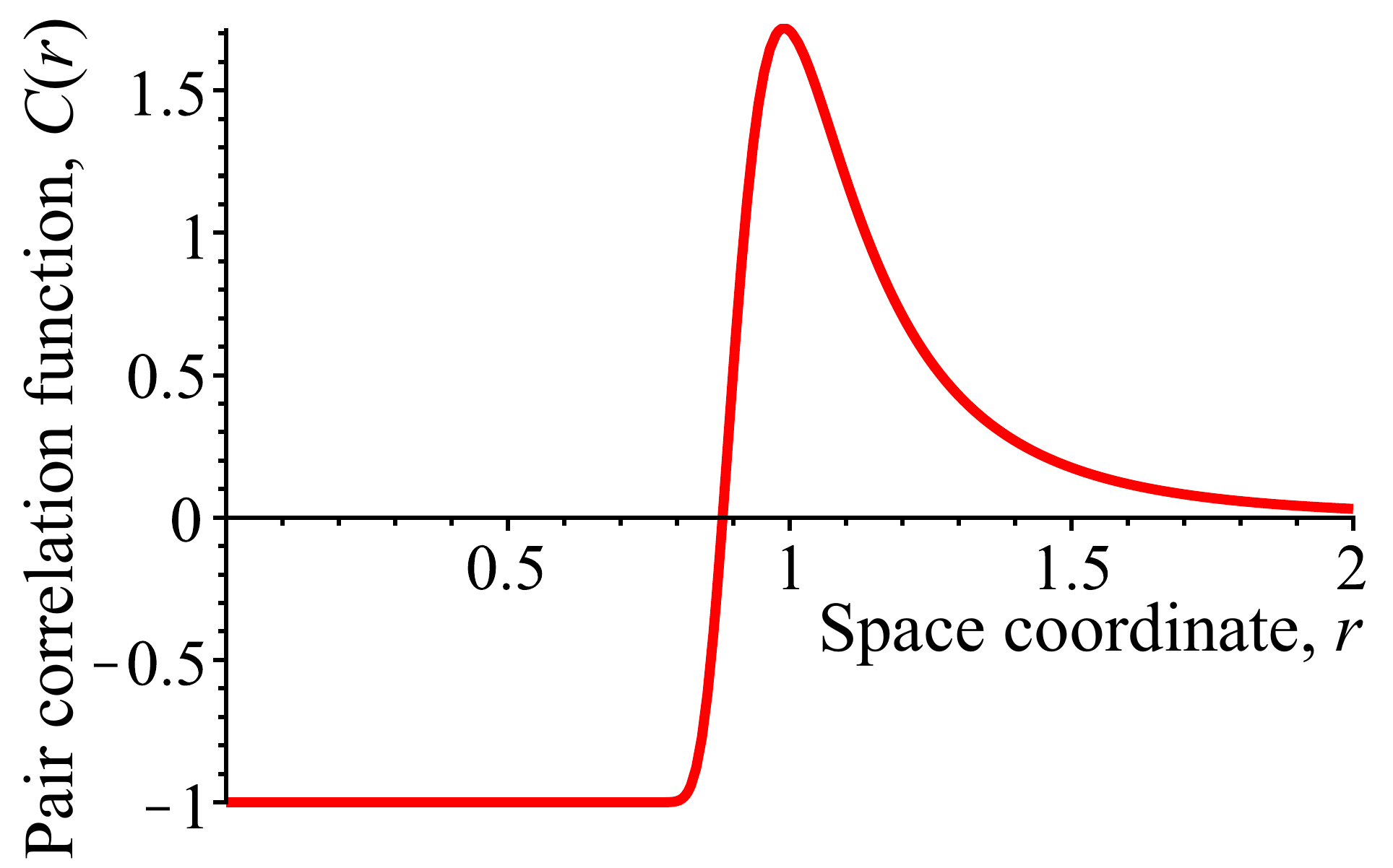}\\
(b)    \includegraphics[width=0.92\columnwidth]{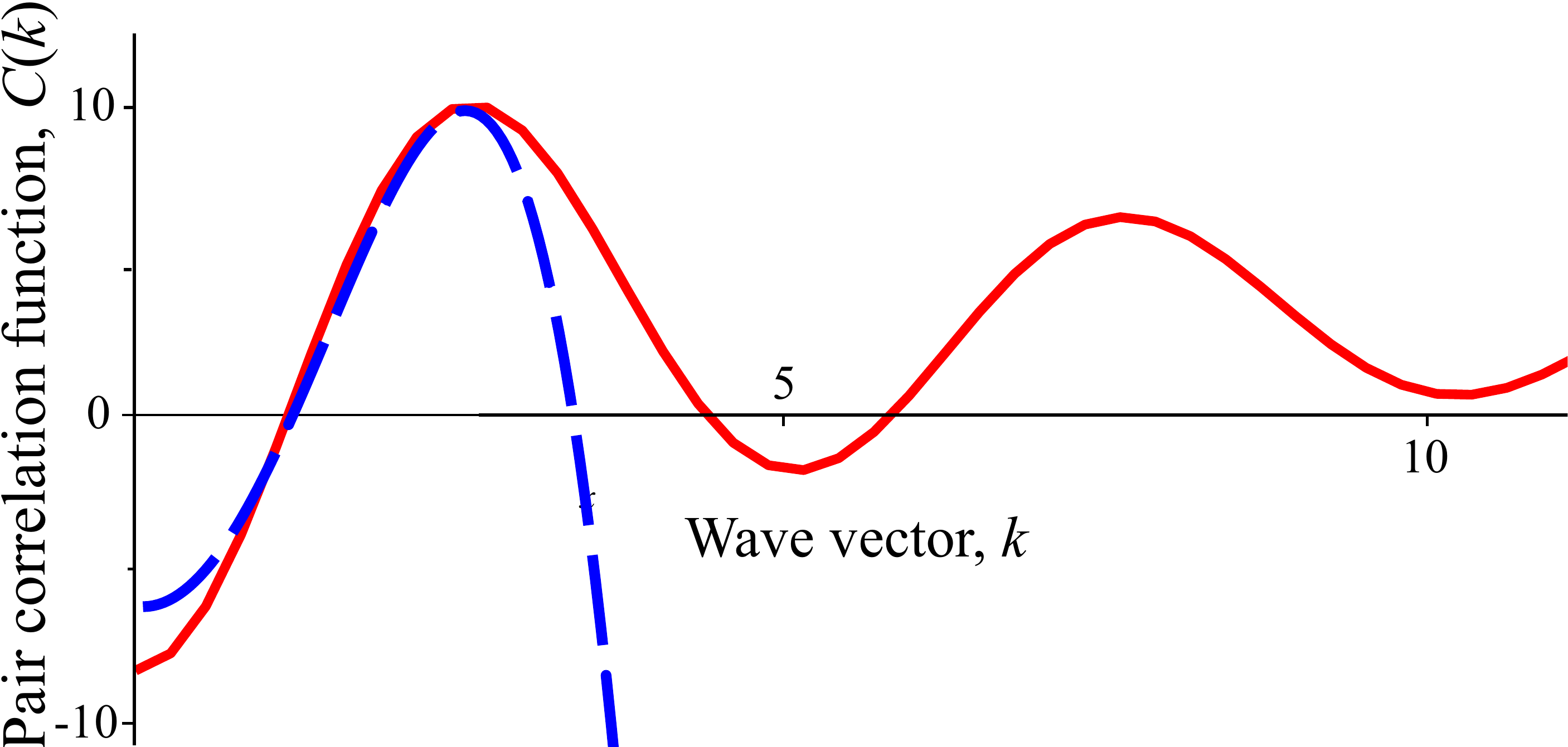}
  \caption{\label{ckfig} Pair correlation functions constructed for (a)  LJ-potential in direct space; (b)  LJ-potential (solid red) and classical PFC (dashed blue) in reciprocal space.}
\end{figure}
The RY approximation of non-local part of the free energy describing freezing transitions Eq.~(\ref{eq:pfc_der_1})  inputs the pair correlation function of fluid. One can obtain this pair-correlation function in the analytical form from the liquid integral equation theory \cite{Hansen2013}. PPractically, for a case on the non-negligible interactions one can assume a deviation from an ideal gas in a form of virial equation of state which was previously introduced in \cite{Lowen2010,bib:adv2012}. The second-order virial coefficient, which has a physical meaning of direct pair correlation function, can be analytically approximated for an arbitrary pair interaction potential  leading to the virial expression \cite{VanRoij1995,Hansen2013}:
\begin{equation}
{C}_{2}({\bf {r}}-{\bf {r}}')=\exp\left(-\frac{U({\bf {r}}-{\bf {r}}')}{k_B T}\right) - 1 
\label{VirialCorr}
\end{equation}
The correspondent distribution function becomes asymptotically exact to the Boltzmann factor of the pair potential in the low density limit \cite{Hansen2013}.
In present paper we propose a simple Lennard-Jones (LJ) potential \cite{Jones1924} for benchmarking the possibility of working with the arbitrary interacting potentials in PFC model:
\begin{equation}
U({\bf {r}}-{\bf {r}}')=\varepsilon_U \left(  \left( \frac{r_m}{{\bf {r}}-{\bf {r}}'} \right)^{12} -  2\left( \frac{r_m}{{\bf {r}}-{\bf {r}}'} \right)^6 \right).
\label{LJpotential}
\end{equation}
This potential includes the long-range attraction as well as repulsive component. The classical LJ potential allows solidification to different crystalline structures \cite{Parrinello1980,Finnis1984} from a LJ-liquid. The possible phase diagram is broader than simple structural hard-sphere solidification, which lays underline the classical-PFC approximation~\cite{rama79}. The similar idea for description of binary LJ-mixtures using the cDFT is utilized with the different approximations of short- and long-rage interactions~\cite{Rick1989}. It is important to point out, that LJ-liquid naturally freezes into an FCC solid in three dimensions \cite{Ohnesorge1994a}, which differs from a hard-sphere-like PFC-models resulting BCC-crystal \cite{Alexander1978}.
In Fig.~\ref{ckfig}(a) the direct space pair correlation function  obtained for LJ-potential Eq.~(\ref{LJpotential}) with virial expression Eq.~(\ref{VirialCorr}) is presented, parameters are $\varepsilon_U=1$, $r_m=1$, $k_B T=1$.
Now we transfer to the reciprocal space for correlation functions to compare the $C(k)$ for one-mode PFC  Eq.~(\ref{ckr}) and LJ with the virial approximation, see Fig.~\ref{ckfig}(b). As one can see the PFC model reproduces the first peak of pair correlation function $C(k)$ . In  more complex case $C(k)$ can be approximated with  the  rational function fit \cite{Pisutha-Arnond2013} to carefully reproduce the higher modes. 

\subsection{Dynamical equation}
The PFC and XPFC conserved dynamic equations for atomic density $n$ are described with \cite{elder21,bib:pe}:
\begin{equation}
\frac{\partial n}{\partial t}=\nabla^2 \mu(n), \quad \quad \mu(n) = \frac{\delta F(n)}{\delta n},
\label{dyn}
\end{equation}
where in classical one-mode PFC $ \mu (n) $ is a chemical potential defined by the functional derivative (Gateaux) of the free energy~(\ref{freen}),
\begin{equation}
\mu=\Delta B_0 n - a n^2 + v n^3 + B_0^x (q_{S0}^2+\nabla^2)^2 n.
\label{mumumu1}
\end{equation}
or the sum of two terms:
\begin{equation}
\mu=\frac{\delta F_{id}(n)}{\delta n} + \frac{\delta F_{ex}(n)}{\delta n}.
\label{mumumu}
\end{equation}

To find the dynamical equation for XPFC model we consider the convolution integral of excess part of the XPFC free energy Eq.~(\ref{eq:pfc_der_1}). This functional term $F_{ex}(n)$ enters the functional derivative (Gateaux) for the dynamical PFC equation.
Let one introduce the test function $\theta({\bf r})$ and a small $\epsilon \rightarrow 0$ so the Gateaux variation for this convolution will be:
\begin{align}
&\left. \frac{d}{d\epsilon}F_{ex}(n\!+\!\epsilon \theta)\right|_{\epsilon=0}\!\!\!\!\!=\\
&\!\!\left. \frac{d}{d\epsilon}\! \int \!\!\! d{\bf r}\!\! \int \!\! (n({\bf r})\!+\!\epsilon \theta({\bf r}) ) {C}  ( {\bf r}\!-\!{\bf r}') (n({\bf r}')\!\!+\!\epsilon \theta({\bf r}'))  d {{\bf r}'} \right|_{\epsilon =0}\!\!\!\!\!=\nonumber
\end{align}
\begin{align}
=&
 \frac{d}{d\epsilon}\! \int \!\!  d{\bf r} \int \!\! d {{\bf r}'} {C}  ({\bf r} - {\bf r}')  \times \\
&\times  (n({\bf r})\epsilon \theta({\bf r}') + n({\bf r}')\epsilon \theta({\bf r}) + \epsilon \theta({\bf r})\epsilon \theta({\bf r}') ) \Big|_{\epsilon =0}.\nonumber
\end{align}
Here term $\epsilon \theta({\bf r})\epsilon \theta({\bf r}')$ is neglected by the order of magnitude as a small parameter.  Resulted functional derivative is
\begin{equation}
\label{fex-der-conv}
\frac{\delta F_{ex}(n)}{\delta n}
 =-\frac{1}{2} \left( \int \!\! d {{\bf r}} n({\bf r}) {C}  ( {\bf r} - {\bf r}') + \!\int \!\! d {{\bf r}'} {C}  ( {\bf r} - {\bf r}')    n({\bf r}') \right),
\end{equation}
which allows one to describe the dynamics in the structural-PFC model for the arbitrary pair correlation function. 
Considering the isotropic correlation function   this convolution becomes:
\begin{equation}
\label{eq:isotropic}
\frac{\delta F_{ex}(n)}{\delta n}
 =-\int \!\! d {{\bf {r}}'} {C}  || {\bf {r}} - {\bf {r}}'||  n({\bf {r}}'),
\end{equation}
where ${C}_{2}  || {\bf {r}} - {\bf {r}}'||  $ is the arbitrary pair correlation function in direct space. This function can be obtained in reciprocal space using the convolution theorem  \cite{Seymour2016,Elder2018}. The fitting of this function in reciprocal space is possible thith rational function fitting procedure \cite{Pisutha-Arnond2013}.  Also for calculation in direct space one can use the similar approximation with help of expansion of this convolution integral into two complex parts. In case of analytical form of $C_2$ in direct space one can introduce  the direct convolution integration to the numerical procedure presented in the next section.
Resulted XPFC chemical potential  after substitution of the functional derivative of $F_{ex}$ Eq.~(\ref{eq:isotropic}) to Eq.~(\ref{mumumu})  will be:
\begin{equation}
\mu(n)=(1-\varepsilon) n({\bf r}) + a n({\bf r}) ^2 + v n({\bf r}) ^3 -
\int \!\! d {{\bf r}'} {C}_{2}  ( {\bf r} - {\bf r}') n({\bf r}').
\label{mumumu2}
\end{equation}

\subsection{Numerical implementation}
One can split Eqs.~(\ref{dyn}) with Eq.~(\ref{mumumu2}) reducing the order of spatial derivatives:
\begin{eqnarray}
\left\{
\begin{array}{ll}
\displaystyle\frac{\partial n}{\partial t}=\nabla^2 \mu, \\
 \mu  =  (1-\varepsilon) n + a n^2 + v n^3 -\! \displaystyle\int \!\!  d {{\bf r}'} {C}_{2}  ( \xi) n({\bf r}'),
\end{array}
\right.
\label{num}
\end{eqnarray}
where $\xi=|{\bf r}-{\bf r}'|$. To deal with the solution divergence at $\xi=0$ we use LJ-potential Eq.~(\ref{LJpotential}) in the form:
\begin{equation}
U(\xi)=\varepsilon_U \left(  \left( \frac{r_m}{\xi+K} \right)^{12} -  2\left( \frac{r_m}{\xi+K} \right)^6 \right)
\label{LJpotential2}
\end{equation}
where $K=0.01$ is a small shift constant. The system of equations~(\ref{num}) has been solved numerically in one- and two-dimensions in direct space using a direct solver PARDISO for the finite element method with the linear-$C^1$ Lagrange elements utilizing the COMSOL Multiphysics Software~\cite{comsol6} on two-processor AMD Epyc~7302-based computer with 1024 Gb of RAM and 64 cores. To perform convolution kernel integration we use a COMSOL realization of moving-boundary integrals \emph{intop()} with $\xi=x-dest(x)$ (example given in one-dimensional case), where  operator \emph{dest()} correspondent to the moving boundary. To reach the convergence the \emph{nojac()} operator for the convolution integral has been used. This modification omit the nonlinear Jacobian contribution from the current step and use the approximation from the previous one. Thus the size of adaptive time step has been limited by $\delta t=10^{-3}$.
The initial density field was set as $n_0$, crystallization was initiated by introducing the single disturbance at the edge of the domain. The two-dimensional computational domain consists of $L_x\times L_y=10 \times 12$ dimensionless units with maximum triangle grid size $\ell=0.3$; the one-dimensional domain includes $L_x=50$, $\ell=0.1$. The periodic boundary conditions were introduced.

\section{Solutions of one-dimensional XPFC equation}
\begin{figure}
\centering
 \includegraphics[width=0.99\columnwidth]{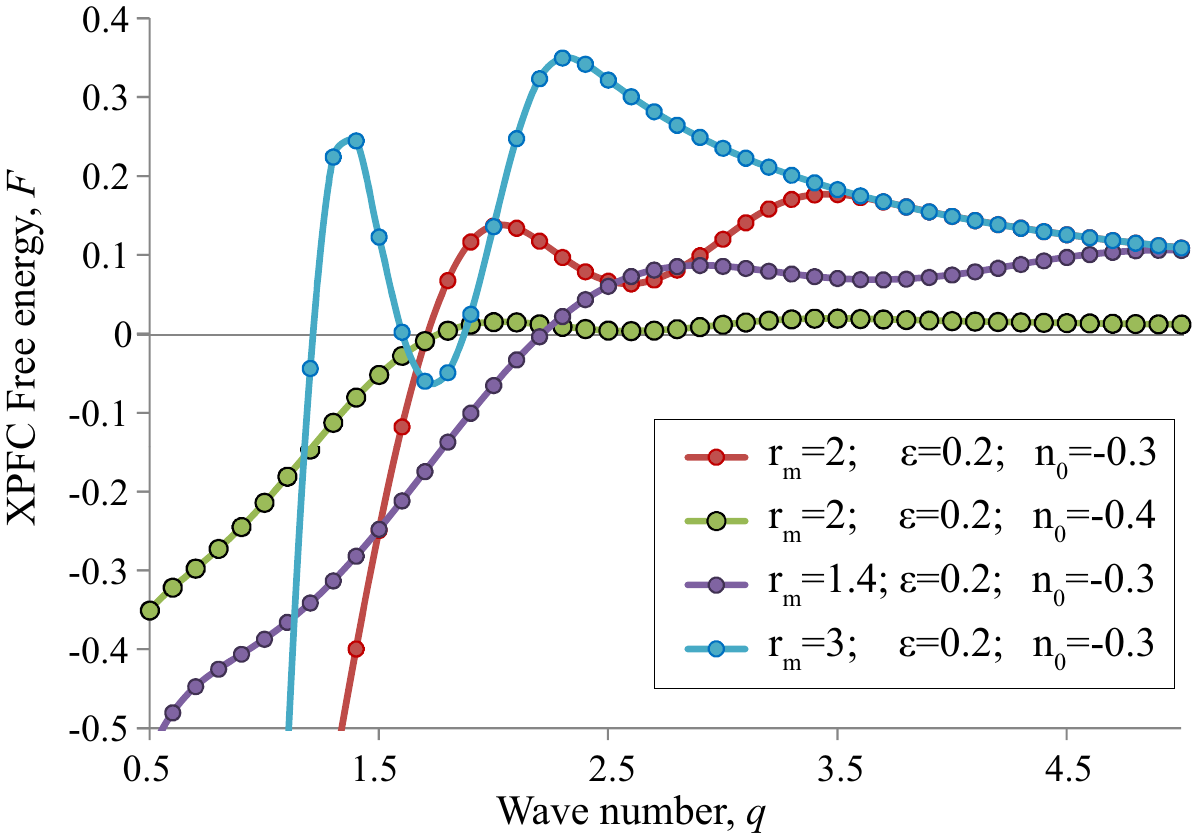} 
  \caption{\label{qmin} Free energy profiles for the periodical solution Eq.~(\ref{cossol}) as a function of wave number $q$ for the one-dimensional XPFC Eq.~(\ref{eq:pfc_der_1}) with LJ-potential obtained for various parameters of LJ-distance $r_m$, driving force $\varepsilon$, density $n_0$.}
\end{figure}

The search for the equilibrium lattice parameter $\lambda$ in the PFC models is of considerable interest. 
We performed numerical minimization of the free energy functional integrated in the limits of a single unit cell $0..2\pi/q$, where $q$ is the wave number. After substituting the simple periodical solution
\begin{equation}
n=n_0 + \eta \mathrm{cos}(q x)
\label{cossol}
\end{equation}
to the XPFC free energy Eq.~(\ref{eq:pfc_der_1}) with analytical correlation function Eq.~(\ref{VirialCorr}) for LJ-potential  Eq.~(\ref{LJpotential}) one can plot a free-energy profiles, see Fig.~\ref{qmin}. The minimization of $\eta$ was performed numerically.
We calculated the preferable wave numbers $q$ for different LJ-potential's parameters and found an inverse dependence of $q$ on parameter $r_m$. One can see a presence of distinct minima correspondent to the preferable $q$ for every profile. We found that $q=3.7$ for $r_m=1.4$ ; $q=2.7$ for $r_m=2$; and $q=1.72$ for $r_m=3$. This also coincides to the minimum  of LJ-potentials which equals to  $r=r_m 2^{1/6}$ \cite{Jones1924}.
  The change of the driving force (undercooling) $\varepsilon$ in the region of stable existence of periodic phase  leads only to the change of PFC amplitudes $\eta$ and quantity of $F$ but not to the change of the position of minimum, \emph{e. g.} equilibrium wave number $q$. As a generalized driving force contribution, initial density $n_0$ also affects only on severity and relative  well depth. Reduction of the density $n_0$ followed by the decreased stability of the periodic phase. When the system approaches to the melting line the minimum disappears. 
  
\begin{figure}[!h]
\centering
(a)\includegraphics[width=0.92\columnwidth]{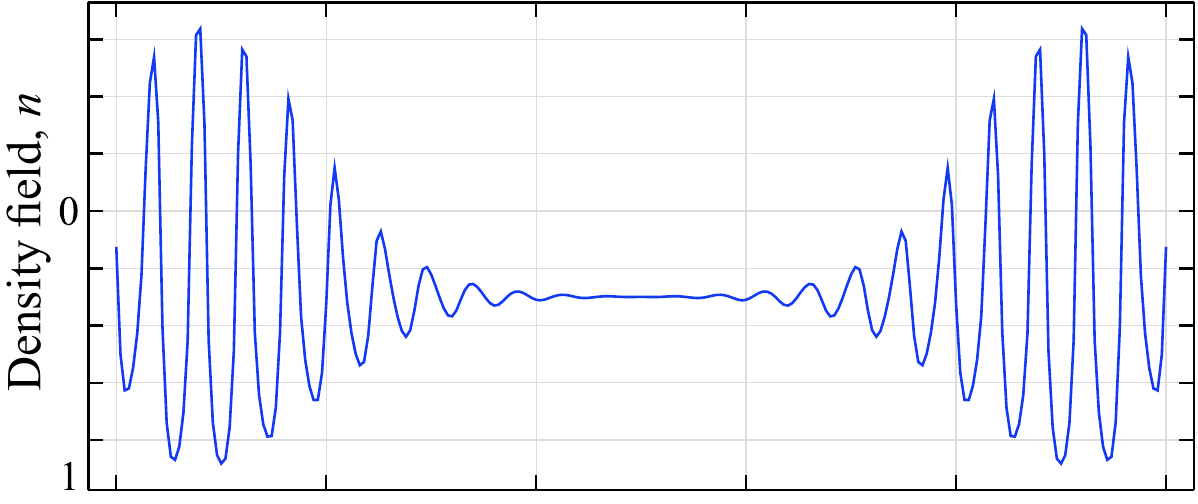}\\
(b)\includegraphics[width=0.92\columnwidth]{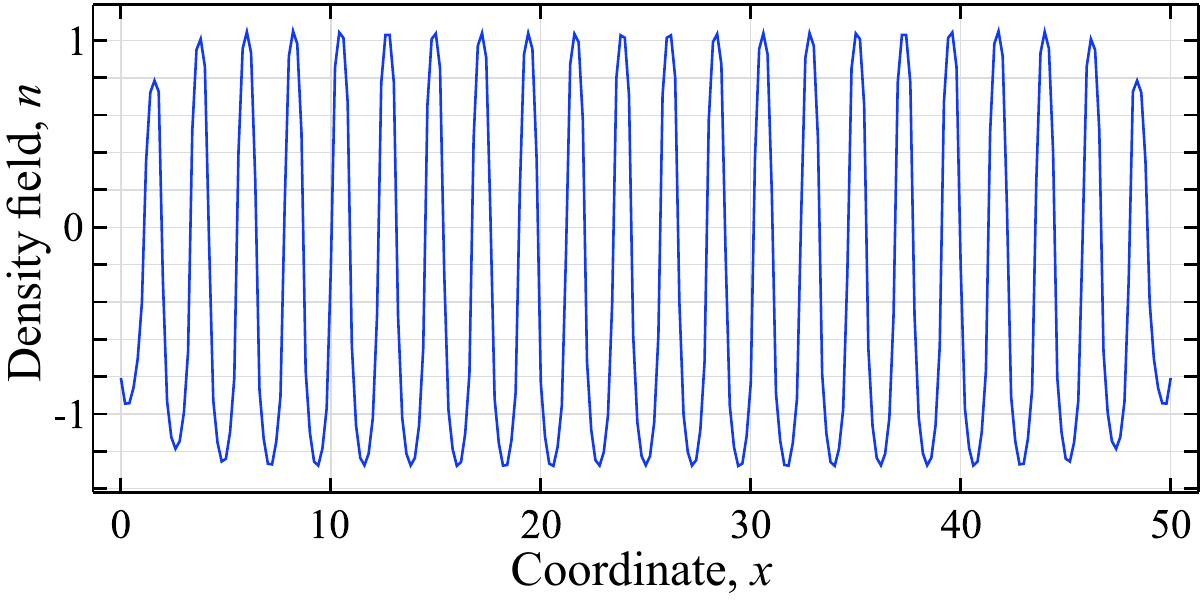}
   \caption{\label{1d} Results of the numerical simulations of the 1D XPFC equation with LJ-potential, parameters are: $r_m=2$, $\varepsilon=0.2$, $n_0=-0.3$. Dynamical distribution of the  density field $n$ and the  phase boundary between periodic lattice and homogeneous phase is presented. Snapshots are given at: (a) $t=0.6$; (b) $t=10$.}
\end{figure}
 
The numerical solutions of  XPFC Eq.~(\ref{num}) for LJ-potential are shown in Fig.~\ref{1d}. The  relatively large interface width is very close to the one shown for the XPFC model with exponential kernel with $\alpha=2$, Ref.~\cite{Greenwood2010}. We found the equilibrium lattice parameter $\lambda=2.3$, which totally coincides with the correspondent wave number $q=2.7=2\pi/\lambda$ obtained by the minimization of $F$ Fig.~\ref{qmin}. The artifacts of the numerical integration procedure is quite noticeable near the edges. Here we see the main problem of the presented implementation of the direct convolution integration. Since the integration is being held for the whole domain  the integration radius is not limited, nevertheless, the integration on boundaries does not give the exact values in the primary peak correlation radius.  Thus the additional development of the numerical implementation is needed. In such case, the reciprocal Fourier spectral method \cite{Pisutha-Arnond2013} looks promising but it also raises the problem of the consideration of the arbitrary boundary conditions and implementation for asymmetrical domains. Besides of that, the real space implementation can improve calculation speed with adaptive mesh refinement; however one cannot expect the unconditional convergence of the numerical scheme in the reciprocal space for the non-rational polynoms including the functions in the form of LJ-potential.  In present work we focused on the implementation of XPFC model with the arbitrary  $C({\bf r})$ kernel with direct finite element scheme in the context of our previous works \cite{Ankudinov2020c,Ankudinov2020}. In addition, our proposed method differs from Helmholtz equations method from \cite{Pisutha-Arnond2013} by direct introduction of the integration of  $C({\bf r})$  over the whole domain.

\section{Solutions of two-dimensional XPFC equation}

\begin{figure}[t]
\centering
(a)\includegraphics[width=0.25\columnwidth]{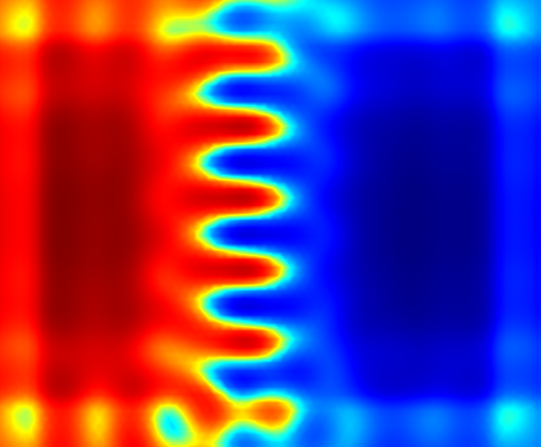} (c)\includegraphics[width=0.25\columnwidth]{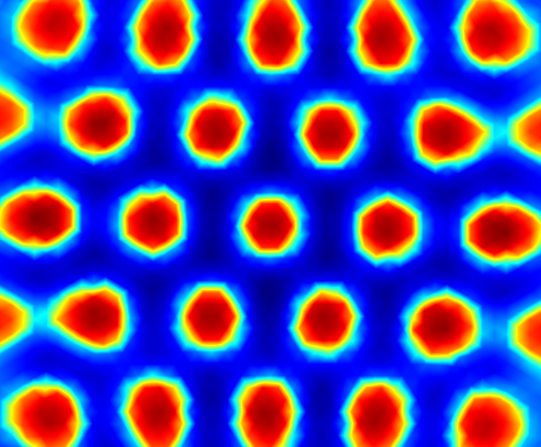} (e)\includegraphics[width=0.25\columnwidth]{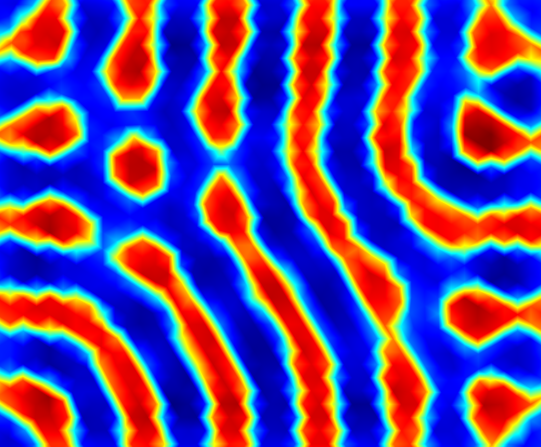} \\
(b)\includegraphics[width=0.25\columnwidth]{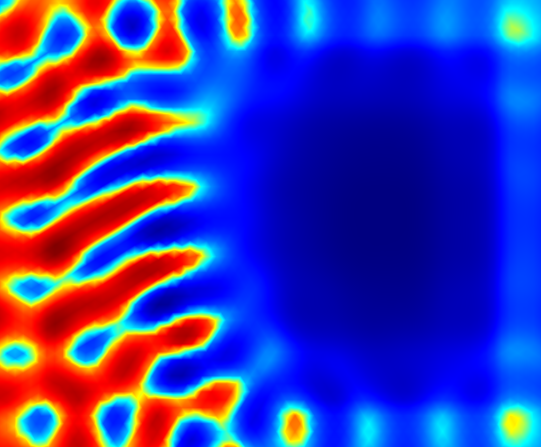} (d)\includegraphics[width=0.25\columnwidth]{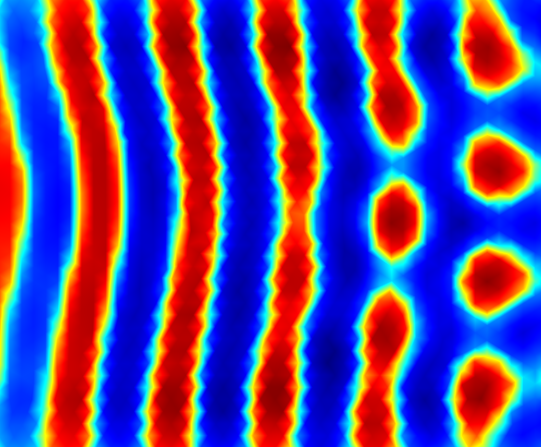} (f)\includegraphics[width=0.25\columnwidth]{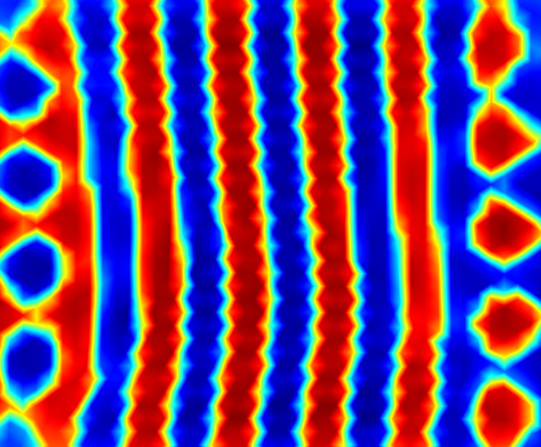}
  \caption{\label{numf} Snapshots of the numerical solutions of 2D XPFC with LJ-potential obtained for different parameters are given at fixed time $t=150$. \\
Column 1, $r_m=1.4$: (a) $\varepsilon=0.1$, $n_0=-0.1$; (b) $\varepsilon=0.5$, $n_0=-0.5$.\\
Column 2, \,\,$r_m=2$:\,\,\, (c) $\varepsilon=0.1$, $n_0=-0.5$; (d) $\varepsilon=0.2$, $n_0=-0.2$.\\
Column 3, \,\,$r_m=3$:\,\,\, (e) $\varepsilon=0.3$, $n_0=-0.5$; (f) \,$\varepsilon=0.4$, $n_0=-0.05$.}
\end{figure}

\begin{figure}
\centering
 \includegraphics[width=0.68\columnwidth]{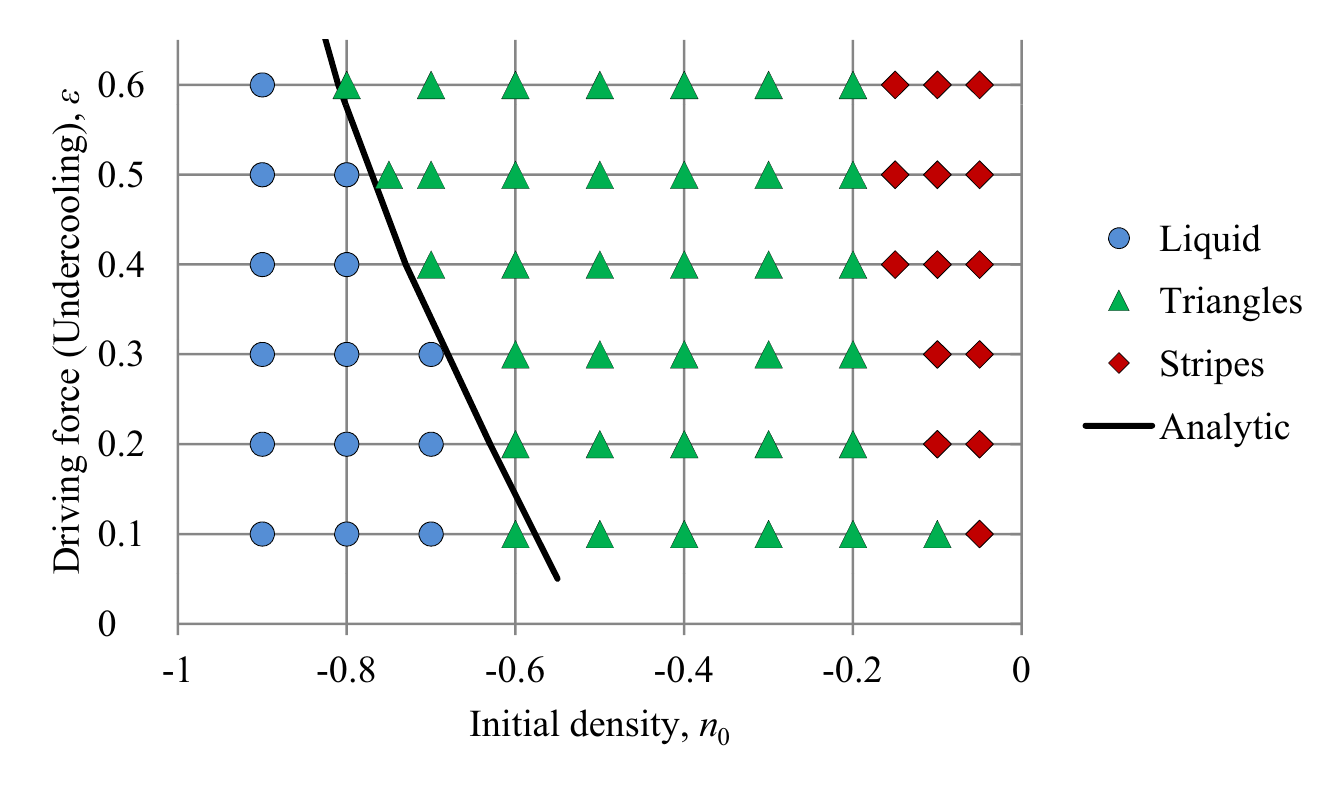} 
  \caption{\label{dia} Numerically obtained phase diagram ``Driving force, $\varepsilon$ -- Initial density, $n_0$'' for the XPFC model with LJ-potential for $r_m=2$. The types of relaxed structures are designated with the correspondent symbols: $\circ$ -- liquid (homogeneous) phase, $\triangle$ -- triangle lattice, $\diamond$ -- striped phase. The analytically obtained melting curve depicted as a solid line.}
\end{figure}

During the numerical simulations of the XPFC-model Eq.~(\ref{num}) with LJ-potential  we obtained a set of various density $n_0$ distribution presented  in Fig.~\ref{numf}. 
For the LJ-distance parameter value $r_m=1.4$ we didn't get any stable triangle structures.  Instead we found a spinodally decomposed homogeneous (liquid) phase with confluent stripe-like phase boundary, Fig.~\ref{numf}(a), and coexistence of stripe-liquid phases Fig.~\ref{numf}(b).
This can be caused by the difficulties in stabilization of the periodic phase at low-range $r_m$, see the correspondent shallow well in Fig.~\ref{qmin}. The other reason is caused by the base property of the PFC models which suppresses the high-frequency harmonics leading to the destruction of high-frequency phase~\cite{ge-2011,Ankudinov2020c}.
Obtained triangle structure is consistent with obtained in PFC models \cite{ankudinov16}, see Fig.~\ref{numf}(c), where the case of $r_m=2$ is considered. The LJ-distance parameter $r_m=2$ was also used during the construction of phase diagram, see Fig.~\ref{dia}. The sample of coexistence of triangle and striped phase for the case of $r_m=2$ is shown in Fig.~\ref{numf}(d). The sample snapshots for  $r_m=3$ is presented in Fig.~\ref{numf}(e) and Fig.~\ref{numf}(f). Here the mixed triangle-striped and striped-inverted trianlge are shown respectively. The boundary artifacts caused by integration are  especially noticeable here. The lattice parameters obtained in this simulations can be compared to the equilibrium ones from the free energy minimization: for $r_m=2$ in two dimensional case we got for triangles $\lambda_\triangle=2.32$, and for stripes $\lambda_\diamond=2.04$; for $r_m=3$ triangles $\lambda_\triangle=3.57$, stripes $\lambda_\diamond=3.02$. 
One can find a difference between $\lambda$ for triangle and striped structures, meanwhile  $\lambda_\triangle$ matches the related $\lambda$ of simple 1D periodical lattice, see Fig.~\ref{qmin}, and equilibrium $\lambda$ obtained from LJ-distance parameter $r_m$. The triangle crystal is derived as a preferable structure for the crystallization  in the theory of  freezing from the isotropic liquids, and in 2D it can be threaten as a simplest possible crystal~\cite{Alexander1978}. The triangle symmetry is analogue to the simplest possible periodical structure in 1D. For 3D case such simplest and most preferable structure is BCC~\cite{Alexander1978}. The lattice matching is related to this fact as soon as the liquid in XPFC model considered as a constant isotropic phase.

\onecolumngrid

\twocolumngrid

With the numerical simulations we obtained the phase diagram of two-dimensional structures for XPFC model with LJ-potentials. Such stable structures qualitatively coincided with the classical PFC (see diagram in \cite{ankudinov16}). We obtain three regions of existence of triangles, stripes and liquid (homogeneous) phases. Results of performed numerical experiments are depicted with the correspondent symbols  in Fig.~\ref{dia}. The analytical curve has been obtained using the thermodynamical method~\cite{ankudinov16} with the numerical minimization of triangle's  and homogeneous free energies and Maxwell area rule. 
The several boundary marks could correspond to the coexistence regions of the adjacent phases such as presented in Fig.~\ref{numf}(d). Obtained phase diagram qualitatively corresponds to the high-temperature region of 2D LJ diagram obtained with molecular dynamics \cite{Barker1981}. Although the striped phase is not observed in LJ diagram. The striped phase here becoming the global free energy minimum among other structures close to the $n=0$. This behavior originates from the Taylor expansion of the local part $F_{id}$ of the free energy, when the term $\textrm{ln}(1+n)$ replaced by a polynomial with two roots \cite{Archer2019}.
The regions of stable structure's existence depends, among other things, on $r_m$ parameter.

\section{Conclusions}
In present work we introduce the modification of the structural phase-field crystal (XPFC) model for Lennard-Jones pair interaction potential. The pair interactions is approximated using the second order term of the virial expansion and thus with the analytical form of the pair correlation function. We studied the presented model with simple LJ potential in one- and two-dimensional cases. The numerical implementation is also presented and discussed. The crystallization from the undercooled homogeneous liquid to the solid crystalline phase was carried out. In one-dimensional case the equilibrium lattice parameters were calculated and compared to the LJ-distance parameters and numerical solutions. The two-dimensional numerical simulations was performed and resulted liquid, triangle, striped phases and their mixtures. 
With the obtained numerical data the lattice parameters for each structure was calculated. 
The lattice parameter's values of the two-dimensional triangle  and one-dimensional periodical structure are exactly matched to each other and to the correspondent LJ-distance parameters. 
The set of obtained numerically structures and the phase diagram by itself qualitatively coincide to the classical PFC model, the form of the melting region is consistent with the high-temperature region of two-dimensional LJ diagram.

\section*{Acknowledgments} We thank N. M. Chtchelkatchev for the valuable discussions on the formulation of the model. This study was financially supported by Russian Science Foundation, project 21-73-00263, https://rscf.ru/project/21-73-00263/.

\providecommand{\newblock}{}


\begin{thebibliography}{10}
\expandafter\ifx\csname url\endcsname\relax
  \def\url#1{{\tt #1}}\fi
\expandafter\ifx\csname urlprefix\endcsname\relax\def\urlprefix{URL }\fi
\providecommand{\eprint}[2][]{\url{#2}}

\bibitem{bib:pe}
Provatas N and Elder K 2010 {\em {Phase-Field Methods in Materials Science and
  Engineering}\/} (Wiley-VCH) ISBN 9783527407477

\bibitem{elder12}
Elder K~R, Katakowski M, Haataja M and Grant M 2002 {\em Phys. Rev. Lett\/}
  {\bf 88} 245701

\bibitem{elder21}
Elder K~R and Grant M 2004 {\em Physical Review E\/} {\bf 70} 51605

\bibitem{Greenwood2011}
Greenwood M, Rottler J and Provatas N 2011 {\em Physical Review E\/} {\bf 83}

\bibitem{Toth2011}
T{\'{o}}th G~I, Pusztai T, Tegze G, T{\'{o}}th G and Gr{\'{a}}n{\'{a}}sy L 2011
  {\em Physical Review Letters\/} {\bf 107} 175702

\bibitem{Landau1996}
Landau L~D, Lifshits E~M and Pitaevskii L~P 1996 {\em {Statistical physics}\/}
  (Butterworth-Heinemann, c1980) ISBN 9780750633727

\bibitem{brazovskii75}
Brazovskii S~A 1975 {\em Journal of Experimental and Theoretical Physics (Rus.
  Zhurnal Eksperimentalnoi I Teoreticheskoi Fiziki)\/} {\bf 68} 175--185 ISSN
  1098-2744

\bibitem{bib:L}
Kats E~I, Lebedev V~V and Muratov A~R 1993 {\em Physics Reports\/} {\bf 228}
  1--91

\bibitem{bib:adv2012}
Emmerich H, L{\"{o}}wen H, Wittkowski R, Gruhn T, T{\'{o}}th G~I, Tegze G and
  Gr{\'{a}}n{\'{a}}sy L 2012 {\em Advances in Physics\/} {\bf 61} 665--743

\bibitem{Granasy2019}
Gr{\'{a}}n{\'{a}}sy L, T{\'{o}}th G~I, Warren J~A, Podmaniczky F, Tegze G,
  R{\'{a}}tkai L and Pusztai T 2019 {\em Progress in Materials Science\/} {\bf
  106} 100569

\bibitem{VanTeeffelen2009}
{Van Teeffelen} S, Backofen R, Voigt A and L{\"{o}}wen H 2009 {\em Physical
  Review E\/} {\bf 79} 051404

\bibitem{GDL}
Galenko P, Danilov D and Lebedev V 2009 {\em Physical Review E\/} {\bf 79}
  51110

\bibitem{granasy}
Tegze G, Gr{\'{a}}n{\'{a}}sy L, T{\'{o}}th G~I, Podmaniczky F, Jaatinen A,
  Ala-Nissila T and Pusztai T 2009 {\em Physical Review Letters\/} {\bf 103}
  035702

\bibitem{Elder2007}
Elder K~R, Provatas N, Berry J, Stefanovic P and Grant M 2007 {\em Physical
  Review B\/} {\bf 75} 64107 ISSN 10980121

\bibitem{Guerdane2018}
Guerdane M and Berghoff M 2018 {\em Physical Review B\/} {\bf 97} 144105

\bibitem{Fallah2013}
Fallah V, Ofori-Opoku N, Stolle J, Provatas N and Esmaeili S 2013 {\em Acta
  Materialia\/} {\bf 61} 3653--3666

\bibitem{Toth2010}
T{\'{o}}th G~I, Tegze G, Pusztai T, T{\'{o}}th G and Gr{\'{a}}n{\'{a}}sy L 2010
  {\em Journal of Physics Condensed Matter\/} {\bf 22} 364101

\bibitem{Tang2011}
Tang S, Backofen R, Wang J, Zhou Y, Voigt A and Yu Y~M 2011 {\em Journal of
  Crystal Growth\/} {\bf 334} 146--152

\bibitem{Podmaniczky2017}
Podmaniczky F, T{\'{o}}th G~I, Tegze G, Pusztai T and Gr{\'{a}}n{\'{a}}sy L
  2017 {\em Journal of Crystal Growth\/} {\bf 457} 24--31

\bibitem{rama79}
Ramakrishnan T~V and Yussouff M 1979 {\em Physical Review B\/} {\bf 19}
  2775--2794

\bibitem{Ryzhov1979}
Ryzhov V~N and Tareyeva E~E 1979 {\em Physics Letters A\/} {\bf 75} 88--90

\bibitem{bib:berry2008}
Berry J, Elder K~R and Grant M 2008 {\em Physical review B\/} {\bf 77} 224114

\bibitem{asadi14}
Asadi E, Zaeem M~A and Baskes M~I 2014 {\em JOM\/} {\bf 66} 429--436

\bibitem{Asadi2015a}
Asadi E and {Asle Zaeem} M 2015 {\em JOM\/} {\bf 67} 186--201

\bibitem{Baker2015}
Baker K~L and Curtin W~A 2015 {\em Physical Review B\/} {\bf 91} 014103

\bibitem{Archer2019}
Archer A~J, Ratliff D~J, Rucklidge A~M and Subramanian P 2019 {\em Physical
  Review E\/} {\bf 100} 022140

\bibitem{ge-swift}
Swift J and Hohenberg P~C 1977 {\em Physical Review A\/} {\bf 15} 319--328 ISSN
  10502947

\bibitem{Ankudinov20202}
Ankudinov V, Starodumov I and Galenko P~K 2021 {\em Mathematical Methods in the
  Applied Sciences\/} {\bf 44} 12129--12138

\bibitem{Greenwood2010}
Greenwood M, Provatas N and Rottler J 2010 {\em Physical Review Letters\/} {\bf
  105} 1--4

\bibitem{Greenwood2011a}
Greenwood M, Ofori-Opoku N, Rottler J and Provatas N 2011 {\em Physical Review
  B\/} {\bf 84} 1--10 ISSN 10980121

\bibitem{Ofori-Opoku2013}
Ofori-Opoku N, Fallah V, Greenwood M, Esmaeili S and Provatas N 2013 {\em
  Physical Review B\/} {\bf 87} 134105

\bibitem{Smith2017}
Smith N and Provatas N 2017 {\em Physical Review Materials\/} {\bf 1} 053407

\bibitem{Chan2015}
Chan V~W, Pisutha-Arnond N and Thornton K 2015 {\em Physical Review E\/} {\bf
  91} 053305

\bibitem{Seymour2016}
Seymour M and Provatas N 2016 {\em Physical Review B\/} {\bf 93} 035447

\bibitem{Elder2018}
Elder K~L, Seymour M, Lee M, Hilke M and Provatas N 2018 {\em Philosophical
  Transactions of the Royal Society A\/} {\bf 376}

\bibitem{mkhonta}
Mkhonta S~K, Elder K~R and Huang Z~F 2013 {\em Physical Review Letters\/} {\bf
  111} 35501

\bibitem{Zapolsky2017}
Zapolsky H, Demange G and Abdank-Kozubski R 2017 {\em Diffusion Foundations\/}
  {\bf 12} 111--126

\bibitem{Ghosh2017}
Ghosh S 2017 {\em Computational Materials Science\/} {\bf 138} 384--391

\bibitem{jaatinen09}
Jaatinen A, Achim C~V, Elder K~R and Ala T 2009 {\em Physical Review E\/} {\bf
  80} 1--10

\bibitem{Ankudinov2020a}
Ankudinov V, Elder K~R and Galenko P~K 2020 {\em Physical Review E\/} {\bf 102}
  062802

\bibitem{Ankudinov2020b}
Ankudinov V 2021 {\em Mathematical Methods in the Applied Sciences\/} {\bf 44}
  12203--12210

\bibitem{Kondo2021}
Kondo R 2021 {\em Physical Review B\/} {\bf 104} 1--14

\bibitem{Pisutha-Arnond2013}
Pisutha-Arnond N, Chan V~W, Iyer M, Gavini V and Thornton K 2013 {\em Physical
  Review E\/} {\bf 87} 1--14

\bibitem{wu07}
Wu K~A and Karma A 2007 {\em Physical Review B\/} {\bf 76} 184107

\bibitem{evans79}
Evans R 1979 {\em Advances in physics\/} {\bf 28} 143--200

\bibitem{singh91}
Singh Y 1991 {\em Physics Reports-Review Section of Physics Letters\/} {\bf
  207} 351--444

\bibitem{Evans1992}
Evans D~J and Morriss G 2008 {\em {Statistical mechanics of nonequilibrium
  liquids, second edition}\/} vol 9780521857 (ANU Press) ISBN 9780511535307

\bibitem{Hansen2013}
Hansen J~P and McDonald I~R 2013 {\em {Theory of Simple Liquids}\/} (Elsevier
  Academic Press) ISBN 0080571018

\bibitem{asadi15}
Asadi E and {Asle Zaeem} M 2015 {\em Computational Materials Science\/} {\bf
  105} 110--113

\bibitem{zaeem16}
Emdadi A, {Asle Zaeem} M and Asadi E 2016 {\em Computational Materials
  Science\/} {\bf 123} 139--147

\bibitem{Lowen2010}
L{\"{o}}wen H 2010 {\em Journal of Physics: Condensed Matter\/} {\bf 22} 364105

\bibitem{VanRoij1995}
{Van Roij} R, Bolhuis P, Mulder B and Frenkel D 1995 {\em Physical Review E\/}
  {\bf 52} R1277--R1280

\bibitem{Jones1924}
Jones J~E 1924 {\em Proceedings of the Royal Society A\/} {\bf 106} 463--477

\bibitem{Parrinello1980}
Parrinello M and Rahman A 1980 {\em Physical Review Letters\/} {\bf 45}
  1196--1199

\bibitem{Finnis1984}
Finnis M~W and Sinclair J~E 1984 {\em Philosophical Magazine A\/} {\bf 50}
  45--55

\bibitem{Rick1989}
Rick S~W and Haymet A~D 1989 {\em The Journal of Chemical Physics\/} {\bf 90}
  1188--1199

\bibitem{Ohnesorge1994a}
Ohnesorge R, L{\"{o}}wen H and Wagner H 1994 {\em Physical Review E\/} {\bf 50}
  4801--4809

\bibitem{Alexander1978}
Alexander S and McTague J 1978 {\em Physical Review Letters\/} {\bf 41}
  702--705

\bibitem{comsol6}
COMSOL A~B 2022 {\em {www.comsol.com, COMSOL Multiphysics{\textregistered} v.
  6.0.}, Stockholm, Sweden\/}

\bibitem{Ankudinov2020c}
Ankudinov V, Starodumov I, Kryuchkov N~P, Yakovlev E~V, Yurchenko S~O and
  Galenko P~K 2021 {\em Mathematical Methods in the Applied Sciences\/} {\bf
  44} 12185--12193

\bibitem{Ankudinov2020}
Ankudinov V and Galenko P~K 2020 {\em Journal of Crystal Growth\/} {\bf 539}
  125608

\bibitem{ge-2011}
Galenko P~K and Elder K~R 2011 {\em Physical Review B\/} {\bf 83} 64113

\bibitem{ankudinov16}
Ankudinov V, Galenko P~K, Kropotin N~V and Krivilyov M~D 2016 {\em Journal of
  Experimental and Theoretical Physics\/} {\bf 122} 298--309

\bibitem{Barker1981}
Barker J~A, Henderson D and Abraham F~F 1981 {\em Physica A\/} {\bf 106}
  226--238

\end{thebibliography}
\end{document}